\documentclass[sigconf]{acmart}
\usepackage{booktabs} % For formal tables
\usepackage{xspace}

\usepackage{multirow} % for tables
\usepackage{array} % for tables, required for tabularx
\usepackage{tabularx} % for tables
\usepackage{enumitem}

\usepackage{algorithm,algpseudocode}
\usepackage{amsmath}
\usepackage{amsfonts}

\usepackage{stmaryrd}

\graphicspath{{Figures/}}

\setlist{leftmargin=5.5mm}

\newcommand{\ie}{\emph{i.e.,}\xspace}
\newcommand{\eg}{\emph{e.g.,}\xspace}

\newcommand{\paratitle}[1]{\vspace{1ex}\noindent \textbf{#1}}

\def\BibTeX{{\rm B\kern-.05em{\sc i\kern-.025em b}\kern-.08emT\kern-.1667em\lower.7ex\hbox{E}\kern-.125emX}}

\begin{document}

\title{Towards Reducing Manual Workload in Technology-Assisted Reviews: Estimating Ranking Performance}

\author{Grace E. Lee}
\authornote{This work has been done when she was a PhD candidate.}
\affiliation{%
  \institution{School of Computer Science and Engineering, Nanyang Technological University, Singapore}
}
\email{leee0020@e.ntu.edu.sg}
\author{Aixin Sun}
\affiliation{%
  \institution{School of Computer Science and Engineering, Nanyang Technological University, Singapore}
}
\email{axsun@ntu.edu.sg}

\begin{abstract}
Conducting a systematic review (SR) is comprised of multiple tasks: (i) collect documents (studies) that are likely to be relevant from digital libraries (\eg PubMed), (ii) manually read and label the documents as relevant or irrelevant, (iii) extract information from the relevant studies, and (iv) analyze and synthesize the information and derive a conclusion of SR. When researchers label studies, they can screen ranked documents where relevant documents are higher than irrelevant ones. This practice, known as screening prioritization (\ie document ranking approach), speeds up the process of conducting a SR as the documents labelled as relevant can move to the next tasks earlier. However, the approach is limited in reducing the manual workload because the total number of documents to screen remains the same. Towards reducing the manual workload in the screening process, we investigate the quality of document ranking of SR. This can signal researchers whereabouts in the ranking relevant studies are located and let them decide where to stop the screening. After extensive analysis on SR document rankings from different ranking models, we hypothesize `topic broadness' as a factor that affects the ranking quality of SR. Finally, we propose a measure that estimates the topic broadness and demonstrate that the proposed measure is a simple yet effective method to predict the qualities of document rankings for SRs.

\end{abstract}


\maketitle

%==================================
\section{Introduction}
\label{sec:Introduction}
%==================================

Systematic reviews are a type of literature review that summarizes all existing studies under a specific clinical question and provides a conclusive answer. As answers presented in SRs are used to inform clinical professionals and often policy makers, it is critical for SR researchers to include all existing studies and derive reliable answers. So conducting a systematic review is comprised of multiple tasks which are designed to achieve total recall as much as possible.

First, given a clinical question and details for determining relevant studies, researchers collect studies that are likely to be relevant to the question from digital libraries (\eg PubMed and MEDLINE) using keyword based Boolean queries. Finding of relevant documents are conducted on collected documents, this first procedure tends to be low precision but aims for high/total recall, resulting in a large document collection. Second, researchers read and label the documents as relevant or irrelevant based on the relevance details. As it is manual process, and there are a large number of document collected (typically several thousands of documents), this step is considered as the most time-consuming task in conducting a SR. This process alone takes weeks to months to complete. Next, researchers extract information such as method and results from identified relevant documents. Note that irrelevant documents are discarded at this point, and only relevant documents are used in the extraction stage. Lastly, researchers analyze and synthesize the results from multiple relevant studies and derive a conclusion for the clinical question they have. We explain each step with more details in Section~\ref{sec:Related_Work}.

There has been active research on improving the expensive screening process, and they can be categorized into two. One stream of work is an automatic screening approach using classification models. Models conduct the screening instead of researchers and identify relevant studies. But a critical shortcoming of this approach is that it is likely to jeopardize the important goal in SRs by making false negative predictions according to the current development. Alternatively, some studies suggest to use models as one of human researchers, so that relevance labels for studies are decided by models and researchers together~\cite{SR_survey}.

The other stream of work is to help researchers during the manual screening by providing the rankings of studies, known as screening prioritization (\ie document ranking). When researchers screen ranked documents where relevant studies are at the top, they will identify relevant ones earlier and move relevant ones the following extraction step. In the meantime, the rest (low-ranking) documents are still under the screening step so the screening and the extraction are conducted simultaneously. A representative work in the screening prioritization is Seed-driven Document Ranking (SDR)~\cite{Seed_driven_SDR}. SDR uses a known relevant document (seed) as a query, and it showed the state-of-the-art performance. Besides, a survey reported that currently the screening prioritization is the most promising approach to be adopted in practice for improving the screening process~\cite{SR_survey}.

However, while the screening prioritization speeds up the process of conducting a SR with early identified relevant studies, it is limited in reducing the manual workload. The total number of documents that researchers deal with during the screening remains the same. In this work, towards reducing the manual work in the screening, we investigate the quality of document ranking. Ranking quality information can be presented to researchers together with prioritized studies at the beginning of screening. ``The estimated position of the last relevant document is in Rank 30'' or ``Other SRs whose ranking qualities are similar to the given SR had all relevant documents within Top 30.'' is an example we can provide to researchers from the ranking quality estimation. This will signal researchers whereabouts relevant documents are ranked and more importantly, let them decide where to stop the screening.

Ranking quality is intertwined with a ranking model: depending on a model's performance, ranking quality varies. We first make analysis on rankings obtained from multiple well-known and high-performing models in the SDR setting. The findings are different ranking models produce different ranking qualities for a SR, but the changes are not significant. Also, when multiple SRs are compared each other in terms of their ranking qualities, some SRs tend to have the higher ranking qualities than other SRs, and this pattern is consistent across different ranking models. Our findings imply there can be a defining characteristic within SRs which underpins their ranking qualities in spite of different ranking models.

In this work, we hypothesize the \textit{topic broadness} of SR as such factor that affects its ranking quality. The rationale is this: when a SR has a broad topic, relevant documents can be dissimilar. It makes more difficult to find all relevant documents, and at the same time easier to miss out less similar relevant studies in the ranking. Thus, SRs with broad topic tend to have low ranking quality (\ie high ranking difficulty or low ranking performance from a model's perspective).

To this end, we propose a simple measure that estimates the topic broadness of SR and verify its effectiveness on CLEF eHealth 2018 Task 2 TAR dataset. Besides, we conduct extensive experiments with well-established query performance prediction methods as alternative SR ranking quality estimators. Our experiment results demonstrate that the topic broadness of SR is a simple and effective indicator for its ranking quality.

Finally, we summarize our major contributions below:
\begin{itemize}
    \item {We make thorough analysis on ranking quality of SRs with multiple ranking models. Our observations include that ranking quality changes by ranking models are small. Moreover, some SRs consistently have the higher ranking quality than other SRs, and this pattern is observed across multiple ranking models.}
    \item {We investigate the topic broadness of SR as a factor that affects the ranking quality. We propose a measure that quantifies the topic broadness. The evaluation shows that the topic broadness is an effective indicator for the ranking quality of SR.}
    \item {Ranking quality information informs researchers whereabouts relevant documents are ranked, and eventually it lets them decide when to stop it. Our work sheds light on reducing the manual work on top of prioritized studies. }
\end{itemize}

%==================================
\section{Related Work}
\label{sec:Related_Work}
%==================================

\subsection{Systematic Reviews}
In evidence-based medicine, a systematic review (SR) answers a clinical question based on existing literature. Answers in SRs inform medical professionals and policy makers so it is very critical to derive reliable answers by including \textit{all} existing relevant studies in SRs. However, identifying all relevant documents is a non-trivial task, as it is unknown that how many relevant documents exist, and where and how they are stored, to name a few challenges. To reach the goal systematically, conducting a SR is comprised with multiple tasks. In the following paragraphs, we explain each task in details. Readers who are familiar with systematic reviews may skip this subsection.

As a preparation of conducting a SR, researchers first define a clinical question and relevance conditions. A clinical question is equivalent to a topic of SR. An example of clinical question is `How accurately does ultrasound diagnose common bile duct stones, compared to liver function tests?' Once a clinical question is set up,  relevance conditions are also determined, which are detailed criteria for the relevance of studies. When researchers label studies in a later task, this relevance conditions are used. In the preparation step, researchers usually become to know one or two relevant documents.

Once clinical question and relevance conditions are determined, the first task is to collect studies that are possibly relevant documents to the clinical question from digital libraries (\eg PubMed, MEDLINE and EMBASE). SR researchers and information specialists (\ie librarians) formulate keyword queries based on the clinical question and the relevance conditions and issue them at databases for document retrieval. Boolean retrieval is commonly used in this task and documents that match keyword queries are collected. Since the aim is finding all existing relevant documents, the large number of studies are collected.  
For example, a SR has around 6,000 collected documents on average in CLEF 2018 eHealth Task 2 TAR dataset.

Next is screening the collected documents. SR experts manually label every candidate as relevant or irrelevant according to the relevance conditions. When a document satisfies all of the relevance conditions, it is labeled as relevant and otherwise, irrelevant. This screening process is conducted by two sequential sub-processes to ensure to find all relevant studies: (i) abstract screening and (ii) full document screening. Researchers first read the only abstracts of all candidates and find (intermediate) relevant documents. Next, they read the full document content of the intermediate relevant documents and confirm final relevant documents for a given SR. Due to the large number of collected candidates and also the manual and multi-stage screening, the screening task alone takes several weeks to months to complete.

The fourth step is to extract information from the relevant studies, such as patient information, experiment settings, and results. Automating the extraction step is also active research areas in technology-assisted systematic reviews~\cite{dataExtract_survey}. After the extraction finally researchers synthesize the data via statistical analysis and derive an answer for given SR.

\subsection{Improving Screening Process in Technology-Assisted Reviews}
Existing approaches for improving the screening process are mainly divided into two categories. The first category is to automate in the screening process. Predictive models can be trained, and they classify candidate studies as relevant or irrelevant. However, this approach is too early to be used in practice based on current development, because it is likely to miss relevant studies via false negative predictions. Alternatively, some claim that predictive models play a role as a additional SR researcher in the screening process or replace one of SR researchers needed in the screening~\cite{SR_survey}. And the final labels of studies are collectively decided by predictive models and researchers.

The second category is to assist SR researchers using technologies for efficient workflow. One approach in this category is screening prioritization (\ie document ranking approach), which allows relevant documents to be found as early as possible. And they can move to the next tasks, while other studies are in the screening. Another approach in this category is to provide a macro view of candidate documents, such as presenting similar document clusters or key phrases of document clusters in a visual form. Hence, SR researchers screen similar documents of clusters consecutively for the quicker screening.

Among the aforementioned approaches, the screening prioritization is considered the most reliable approach to be used in practice~\citep{SR_survey}. It improves the process of conducting a SR to be efficient, and it does not cause the risk of missing relevant documents unlike the automatic approaches. For the last couple of years, the CLEF eHealth evaluation lab organized data competitions named Technologically Assisted Reviews (TAR). It has further catalysed research on the screening prioritization and also released benchmark datasets~\citep{clef_task,CLEF18_overview,clef_19_overview}.

In this work, we aim to move towards reducing the manual workload on top of screening prioritization. By providing the additional information on ranking quality, researchers can anticipate how long it takes to find all relevant documents before embarking the manual screening. More importantly, they can decide where to stop screening in prioritized studies. Studying when to stop the screening is not new ~\cite{UvAWhenToStop,SheriffieldWhenToStop}. Previous work estimates a stopping point using document labels produced from the manual screening. In other words, it is necessary for researchers to start the screening. Differed from existing work, our work provides the information that researchers can utilize for a stopping point before they start the screening.

\subsection{Query Performance Prediction}
There is a task similar to the ranking quality estimation for SRs. Query performance prediction (QPP)~\cite{preretrievalQPP_survey_CIKM_08,ICIIR09_NQC_query_commitment,ICTIR_2013_significance_test_reviewer2} is a longstanding research problem in Information Retrieval, established from TREC Robust Track in 2004~\citep{TRECRobust2004overview}. The task is to predict the performance of a query given a document corpus and a search system/model. While existing QPP methods can be adopted and used for estimating the ranking quality of SR~\cite{twosides_of_same_coin_QPP_ClusterRank,twosides_of_same_coin_QPP_effectiveness_eval}, there exists multiple points that set the ranking quality estimation apart from the QPP task. For ease of comparison, we will use `ranking quality' and `performance' and also `estimation' and `prediction' interchangeably.

The first difference is an target object for which we estimate ranking quality. In the QPP task, a query is the target object, and the goal is to estimate its performance. On the other hand, in the ranking quality estimation task, a systematic review is the target object, instead of a query. We estimate the ranking quality of SR, and a query is a known relevant document (seed) and it can be any relevant document of SR. Additionally, for the completeness of the study, we also looked into the performance of each relevant document within a SR, same as the QPP task, but relevant documents result in very similar performances, so that estimating query performance is less useful.

The second is a ranking object. In the QPP task, when we estimate the performance of query, a ranking object is a document corpus and it is typically very large (800,000 documents in TREC Robust Track). However, in the ranking quality estimation, a ranking object is candidate documents, which was retrieved from the previous collection step and they are typically 6,000 to 8,000 documents in CLEF eHealth 2018 Task 2 TAR dataset. More importantly, while the ranking object in the QPP is unchanged when we estimate the performance for different queries, the ranking object in the ranking quality estimation changes depending on SRs. In other words, it is a variable that we need to take into accounts when we estimate the ranking quality of SR. Lastly, a document corpus in the QPP is a collection of random documents, but candidate documents in SRs are a collection of similar documents, since they are retrieved via Boolean search.

Although there are the differences in the target object and the ranking objects in the two tasks, we can choose to leverage existing QPP methods and apply it for the ranking quality estimation for SRs. In this work, we test a wide range of QPP methods from pre-retrieval QPP methods~\cite{preretrievalQPP_survey_CIKM_08} to post-retrieval QPP methods~\cite{ICIIR09_NQC_query_commitment,ICTIR_2013_significance_test_reviewer2} and study if they are effective for estimating the ranking quality of SR, despite the differences in the target and ranking objects.

%==================================
\section{Analysis on Ranking Quality of Systematic Review}
\label{sec:observation:analysis}
%%================================

Ranking qualities of SRs are affected by the choice of ranking models. In this section, we study changes of ranking qualities caused by ranking models. We use three promising ranking models that use different aspects of data to compute ranking scores. They have shown high performance in previous literature.

\textbf{BM25} is a classic retrieval function for bag-of-words (BOW) document retrieval.
It is the most well-known simple and effective baseline in retrieval tasks. It computes a document ranking score using the length of document, within-document term frequency, and inverse document frequency in a document collection. Several existing studies have shown that BM25 also provides solid baseline performance in screening prioritization~\citep{clef_task,CLEF18_overview}

Seed-driven Document Ranking (\textbf{SDR}) is a ranking model specialized for screening prioritization. In SDR, documents are represented as a bag-of-medical concepts (BOC). When calculating ranking scores, SDR uses a weight function for medical concepts in a query. The weight function is combined with a probabilistic query likelihood (QL) model. When we transform documents into BOC representations, we follow the procedure described in the original paper~\citep{Seed_driven_SDR}.

Average embedding similarity (\textbf{AES}) is a simple and effective model using word embeddings in document representations~\citep{Seed_driven_SDR}. A document is represented as the average embedding of the words that appear in the document. Different from BOW and BOC representations, word embedding alleviates the common vocabulary mismatch problem. A ranking score for a document is a cosine similarity with a query. We use publicly available word embedding trained by skip-gram word2vec on the MEDLINE/PubMed corpus\footnote{\label{fn:bioNLPwebsite}\url{http://bio.nlplab.org/}}. The dimension of word embedding is 200, and the window size is set to 5. The rest of the parameters are the same as the original word2vec implementation~\citep{word2vec_original}.

In screening prioritization, a query is not formally defined. Among several suggested resources for a query (details in Related Work), we use an example document (seed document) as a query. This decision has been made since using a seed document as a query does not require the additional step of creating queries, and thus the derived ranking results are easily reproducible. We use every relevant document as a query within a SR, and each document ranking is evaluated by average precision (AP). Lastly, we set the average of AP as the ranking quality of SR.

We use CLEF 2018 eHealth Task 2 TAR (CLEF 2018) dataset. The CLEF 2018
dataset consists of 72 SRs and provides document IDs (PMIDs) for each SR’s candidate documents. We obtain document content (title and abstract) from the MEDLINE/PubMed corpus using PMIDs. Every candidate document has two relevance labels from the abstract screening and the full document screening. The relevant labels are binary (relevant or non-relevant). Because the full document labels are the final labels for screening, we use them to evaluate the ranking model. Table~\ref{tab:data_stat} presents the statistics of candidate documents and relevant documents in the dataset. In CLEF 2018 dataset, a few SRs contain zero relevant documents, even though actual SRs do have relevant documents. The reason is that CLEF 2018 dataset is created from PubMed alone, and thus documents listed in other digital libraries are not included. For details of the creation of CLEF 2018 dataset, we refer to \citep{clef_task,CLEF18_overview}.

\begin{table}
\small
\centering
\caption{The numbers of candidate documents and relevant documents of 72 SRs in CLEF 2018 dataset. The number of relevant documents is based on labels by full document screening.  The numbers in median and average are rounded to the nearest integer. }
\label{tab:data_stat}
\begin{tabular}{c|cc}
\toprule
         & Candidates documents & Relevant documents \\
\midrule
Max & 79,786 &117\\
Min & 64 &0\\
Median & 3,424 & 14\\
\midrule
\textbf{Mean} & \textbf{6,346} & \textbf{23}\\
\bottomrule
\end{tabular}
\end{table}

\textbf{Observation 1.} The changes in ranking quality of a SR by ranking models are very small.

For each SR, we calculate the difference of ranking qualities between two ranking models. After creating pairs from the three ranking models, we take the differences of these pairs and average them. The mean ranking quality difference in all SRs is $0.067$  $(\sigma=0.076)$. This shows that despite different ranking functions, the changes of ranking qualities in SRs are minimal.

\textbf{Observation 2.} Given a ranking model, the difference of ranking qualities among SRs is large, and the large difference is common in all three models.

\begin{figure*}
\centering  
\includegraphics[width=1.0\linewidth]{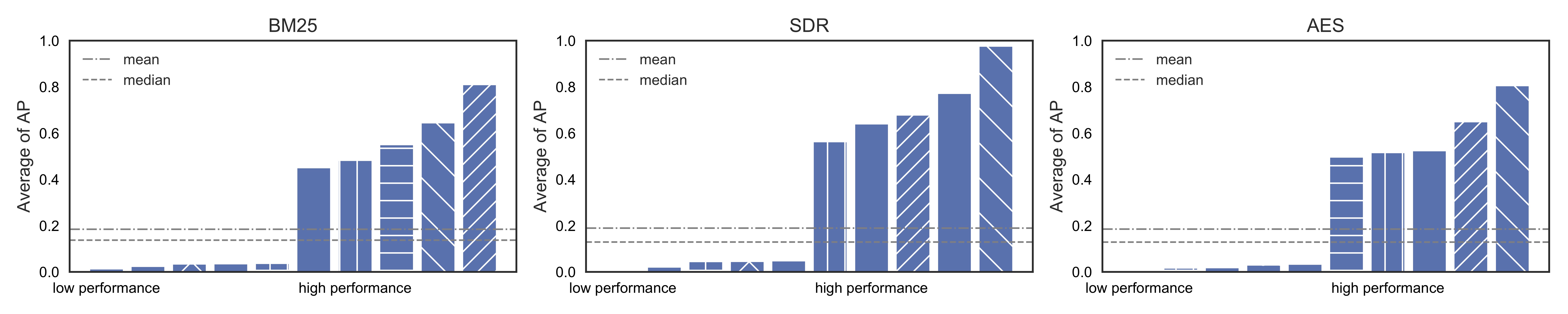}%
\caption{The lowest and the highest 5 ranking qualities of SRs in three ranking models. The pattern on the bars indicates SRs that have the lowest/highest 5 ranking qualities in more than one model.}
\label{fig:observe1:large_perf_variance_3_models}
\end{figure*}

We compare SRs in terms of their ranking qualities given a ranking model. The difference between the highest ranking quality of SR and the lowest ranking quality of SR in BM25 is $0.80$. The gaps between highest and lowest ranking qualities in SDR and AES are $0.97$, and $0.80$, respectively. In all three ranking models, SRs have drastically different ranking qualities. Furthermore, we also present the highest and the lowest 5 ranking qualities within each ranking model in Figure~\ref{fig:observe1:large_perf_variance_3_models}. Observe the substantial gap between two groups of SRs within each model, and the large gaps are consistent in all models.

From these observations, we can come to the conclusion that even though different ranking models are used, when SRs are compared in terms of their rank qualities, their rankings barely change.  We first rank 72 SRs by their ranking qualities using two different ranking models, and then calculate a ranking correlation coefficient (Spearman’s $rho$) between the two lists of SRs. The rank correlation coefficients in pairs of models \{BM25, SDR\}, \{BM25, AES\}, and \{SDR, AES\} are 0.81, 0.77, and 0.75, respectively. Indeed, all three pairs have very strong rank correlations. This results confirm that some SRs consistently have the higher ranking qualities than other SRs despite different ranking models.

Furthermore, the patterned bars in Figure~\ref{fig:observe1:large_perf_variance_3_models} indicate the SRs that present the highest/lowest 5 ranking qualities in more than one model. On the other hand, the bars without patterns are SRs that only appear in a given ranking model, not the other two. The majority of SRs that have the highest/lowest ranking qualities in one model also have the highest/lowest ranking qualities in other models.

This analysis implies that there might be defining factors in SRs that underlie the ranking qualities of SRs. In other words, a SR has characteristics that may affect the document ranking of the SR to be easy (high ranking quality) or difficult (low ranking quality).

%==================================
\section{Ranking Quality Estimation}
\label{sec:proposed_method}
%==================================

In this work, we investigate the \textit{topic broadness} of SR as a factor that influences the ranking quality. The rationale is that the broad topic of SR is likely to lower the ranking quality because its relevant documents are likely to diverse (less similar to each other) and it makes difficult to find them in candidate documents. Furthermore, when a SR has broad topic,  it is not straightforward to accurately encode the relevance in a query and a ranking model, compared to SRs with specific topic. Therefore, SRs with broad topic are likely to have low ranking quality (\ie high ranking difficulty).

We propose a measure that quantifies the topic broadness of SR to estimate the ranking quality. We adopt the same assumption made in \citep{Seed_driven_SDR}: a few example documents (\ie seed documents) are available before the screening step. With the assumption that  two seed documents are available, the proposed measure estimates the topic broadness of SR using the two seed documents and candidate documents.  Specifically, the topic broadness is computed using (i) a similarity between two seed documents and (ii) a similarity between a seed and a candidate documents. Formally, we define topic broadness (TB) as follows.

\begin{equation}\label{Equation:proposed_measure}
	 TB\left(s_1,s_2,C\right) = -\dfrac { 2\cdot  sim\left( s_1,s_2\right) } {\sum _{c\in C}sim\left( s_1,c\right)+\sum _{c\in C}sim\left( s_2,c\right) }
\end{equation}
\begin{equation}\label{Equation:doc_sim_tf}
	 sim(d_1, d_2) = \sum _{w\in d_1,d_2} tf\left( w,d_1\right) + tf\left( w,d_2\right)
\end{equation}
In Equation~\ref{Equation:proposed_measure}, $s_1$ and $s_2$ ($s_1\neq s_2$) denote two seed documents. $c$ is a candidate document in the collection $C$ ($c \in C$). Note that Equation~\ref{Equation:proposed_measure} contains a negative sign. Hence, a great value in nominator leads to small topic broadness (\eg specific topic).
Besides, since $TB$ from  Equation~\ref{Equation:proposed_measure} is the estimated topic broadness of SR, a negative value of $TB$ can be considered as the estimated ranking quality of SR.

We consider the increase of $sim\left( s_1,s_2\right)$, a similarity between two seed documents, makes the topic broadness (ranking quality) of SR decreased (increased). When a SR has broad topic, its relevant documents including $s_1$ and $s_2$ are less similar. The relation of $sim\left( s_1,s_2\right)$ with SR ranking quality is supported by cluster hypothesis: the pairwise similarity of the relevant documents can determine the performance of a search on a particular collection~\citep{ClusterHypothesis_1985}. 

Next, we set the increase of $sim\left( s_*,c\right)$, a similarity between a seed and a candidate document makes the topic broadness (ranking quality) of SR increased (decreased). Since the majority documents are non-relevant documents and only a small fraction of documents in $C$ are relevant documents, $c$ from $C$ is considered a non-relevant document. High similarities between relevant and non-relevant documents affect the low ranking quality of SR as it makes difficult to differentiate relevant  and non-relevant documents. Moreover, in pilot tests we found that a SR tends to have the low rank quality when its number of candidate documents is large. We leverage the size of $C$ by using the sum of $sim(s_*,c)$ in $C$ in $TB$. Lastly, the proposed measure uses the average\footnote{We examine alternative ways (\eg the maximum of the two values) to incorporate two document similarities of each seed and candidate documents, but there was no significant difference.} of two similarities of candidate documents with two seeds.

In Equation~\ref{Equation:proposed_measure}, a similarity between two documents is calculated by Equation~\ref{Equation:doc_sim_tf}.  $tf\left(w,d\right)$ denotes a frequency of word $w$ in document $d$. 
We note that the proposed measure does not incorporate information that are obtained from executing a ranking model such as ranking scores of candidate documents or relevance feedback.

The most similar existing work to the proposed measure is \citep{SIGIR2006what_makes_a_query_difficult_four_distances}.
The authors suggest multiple types of similarities within/across a document collection, relevant documents, and a query, in order to estimate query performance in the QPP task. We evaluate a few similarities proposed in \citep{SIGIR2006what_makes_a_query_difficult_four_distances} that can be applicable to the SR ranking quality estimation.

%==================================
\section{Experiment Setup}
\label{sec:Exp_setup}
%==================================

\paratitle{Selection of Two Seed Documents.}
The proposed measure assumes two relevant documents are available and uses them for estimating the topic broadness of SR. For a given SR, any two documents among relevant documents can be two seed documents. To demonstrate this scenario, we evaluate the proposed measure using all possible pairs of relevant documents in a SR, and report the averaged value over the all pairs. When many pairs are possible in a SR, we use randomly selected 30 pairs among them. Since we use the averaged value, we present the distribution over the estimated values of pairs in the Result and Discussion section and analyze the effect of different seed documents.

\paratitle{Dataset.}
The same CLEF 2018 dataset presented in Analysis section is used.
As the proposed method assumes that two seed documents are available, SRs having less than 2 relevant documents are excluded. Among 72 SRs in CLEF 2018 dataset, 9 SRs have 0 or 1 relevant documents. Total 63 SRs are used in the evaluation.

\paratitle{Baselines.}
We evaluate two sets of baselines: pre-retrieval QPP measures and alternative measures to estimate topic broadness. As discussed in Related Work section, QPP measures are mainly classified into pre-retrieval QPP measures and post-retrieval QPP measures. As the proposed measure does not use any information obtained from ranking models, we set only pre-retrieval QPP measures as baselines for a fair comparison. We use pre-retrieval QPP measures that are frequently evaluated in existing literature~\citep{SIGIR18_QUT_keyword_QPP}: query length (Qlen), a family of IDF-based measures (avgIDF, sumIDF, maxIDF, stdevIDF) \citep{SIGIR02_clarity}, simplified clarity score (\textbf{SCS})~\citep{Jornal_SimplifiedClarityScore}, average inverse collection term frequency (\textbf{avgICTF})~\citep{Jornal_SimplifiedClarityScore}, similarity between a collection and a query (\textbf{SCQ}), and its two variations, normalized SCQ (\textbf{nSCQ}) and maximum SCQ (\textbf{maxSCQ})~\citep{ECIR08_Similarity_Query_Collection}. They incorporate the statistics of terms in a query and the collection of documents in order to estimate the query performance. Due to the lack of space, we explain the details of pre-retrieval QPP baselines in a supplemental material. In the QPP methods, the greater the estimated value indicates the higher the query performance. In this task, we use the performance of query as the ranking quality of SR.

Two estimated values are obtained in the pre-retrieval QPP measures as each of two seed documents can be a query. We examine the average and the maximum of the two QPP estimations as a  final estimation by a pre-retrieval measure. The average value results in slightly better performance than the maximum value but they show no significant difference. Hence, we use the average of the two estimations as the final estimation by the pre-retrieval QPP baselines.

As the second set of baselines, we suggest and examine several alternative measures that quantify the topic broadness of SR using candidate documents:  collSize, collClarity, numClusters, a family of document similarities. 
The rationale is that candidate documents are specifically collected documents for a given SR, so that they might be able to represent the topic broadness of SR.

\textbf{collSize} is the size of document collection (\ie the number of candidate documents). Collection clarity (\textbf{collClarity}) is motivated by Clarity~\citep{SIGIR02_clarity} and simplified clarity score (SCS)~\citep{Jornal_SimplifiedClarityScore}. collClarity  measures a divergence between the candidate collection language model and a corpus (PubMed) language model as follows. 
\begin{equation}
\label{Equation:collection_clarity}
collClarity = \sum _{w\in C}P(w|C)\cdot\log_2  \dfrac {P(w|C)}{P(w|\mathbb{C})}
\end{equation}
where $P(w|C)$ is estimated by maximum likelihood on the candidate collection $C$. 
$P(w|\mathbb{C})$ denotes the corpus language model and we use the uni-gram language model of PubMed corpus\textsuperscript{\ref{fn:bioNLPwebsite}} $\mathbb{C}$. We also evaluate the number of clusters (\textbf{numClusters}) in the collection of candidate documents, proposed in~\citep{SIGIR2006what_makes_a_query_difficult_four_distances}. We first represent each document as the average embedding of all word embeddings included in given document, and then find the best number of clusters a range from 2 to 100 using $k$-means clustering according to silhouette scores.

Lastly, we evaluate a family of pairwise document similarities among candidate documents. 
Likewise, a candidate document is represented as the average embedding of all word embeddings and then we report average (avgDocSim), sum (sumDocSim), maximum (maxDocSim), and standard deviation (stdevDocSim) of the pairwise document similarities.

\paratitle{Ground Truth and Evaluation Measures.} 
We set the ground truth of the ranking quality of SR to its average precision (AP).
AP is an ranking evaluation measure that takes into account both precision and recall components. Specifically, it  averages all precision values at where each relevant document is found in a ranking result. In general, it indicates how good the given ranking result is.
We use the ranking quality by the same  three ranking models (BM25, SDR, and AES) introduced in the previous section, so that the measures are evaluated on the three sets of ground truth.

Our main focus is to estimate the relative ranking qualities of SRs, rather than the actual values of ranking qualities. 
The actual values of ranking qualities change (yet slightly) by ranking models, but the relative ranking qualities among SRs mostly remain unchanged (discussed in the previous section). 
We examine the order of SRs ranked by the estimated ranking qualities, comparing to the order of SRs ranked by ground truth. A non-parametric ranking correlation coefficient (Spearman's $\rho$) is computed between the two lists of SRs. A high absolute value of the coefficient  presents a high accuracy of a measure in estimating the relative ranking qualities of SRs. 
In addition to the main focus, we also report the evaluation results in estimating the actual value of SR ranking quality, using a linear correlation coefficient  (Pearson's $\gamma$). This evaluation setting is similar to that of the QPP task.

%==================================
\section{Results and Discussion}
\label{sec:Results_Discussion}
%==================================

In this section, we discuss the overall evaluation results for the proposed measure and the baselines. Next, we present the detailed analysis on the proposed measure.

Table~\ref{tab:twoseeds_res} shows the overall performance of the proposed measure and the baselines. The best measure is in boldface and the second best measure is underlined. We present the ranking of the proposed method among all evaluated methods and also the performance difference between the proposed method and the best performing baseline in the last row of Table~\ref{tab:twoseeds_res}.
Since the evaluation metrics are correlation coefficients, the +/- signs are the direction of a correlation. We focus on the absolute values of coefficients to compare the effectiveness of measures.
\begin{table*}
\small
\centering
\caption{Evaluation of the proposed measure and the baselines using two seed documents. Spearman's and Pearson's correlation coefficients ($\rho$ and $r$)are computed with the two lists of SR ranked by the estimated values and the actual performance of a given ranking model in the average of AP. The best performance and the second best performance are in boldface and underlined, respectively. Lastly, $p$-values are in parenthesis.}
\label{tab:twoseeds_res}
\begin{tabular}{l|c|c|c|c|c|c}
\toprule
    & \multicolumn{3}{c|}{Spearman's rank correlation coefficient ($p$-value)} &  \multicolumn{3}{c}{Pearson linear correlation coefficient ($p$-value)}    \\
         \midrule
             &   BM25 &                  SDR &                  AES &               BM25 &                  SDR &               AES    \\
\midrule
QLen         & 0.181 (1.62$e$-01)* & 0.341 (7.15$e$-03)* & 0.200 (1.22$e$-01)* & 0.179 (1.66$e$-01)* & 0.369 (3.47$e$-03)*  & 0.180 (1.64$e$-01)*  \\
avgIDF      & -0.148 (2.55$e$-01)* & -0.149 (2.52$e$-01)* & -0.256 (4.58$e$-02)* & -0.297 (2.01$e$-02)* & -0.247 (5.53$e$-02)* & -0.343 (6.73$e$-03)*\\
sumIDF      & 0.062 (6.33$e$-01)* &  0.130 (3.18$e$-01)*   & -0.046 (7.20$e$-01)*  & -0.094 (4.70$e$-01)*& 0.067 (6.07$e$-01)* & -0.154 (2.35$e$-01)* \\
maxIDF       & -0.351 (5.51$e$-03)* & -0.231 (7.40$e$-02)* & -0.477 (1.02$e$-04)* & -0.580 (9.56$e$-07)*& -0.420 (7.44$e$-04)*  & -0.626 (6.72$e$-08)*\\
stdevIDF    & 0.430 (5.49$e$-04)* & 0.464  (1.65$e$-04)*  & 0.492 (5.56$e$-05)*& \textbf{0.701} (3.10$e$-10)& \underline{0.650} (1.40$e$-08)  & \underline{0.703} (2.67$e$-10)* \\
SCS       & -0.057 (6.63$e$-01)* & -0.201 (1.20$e$-01)*  &  -0.191 (1.40$e$-01)* & -0.220 (8.86$e$-02)* & -0.281 (2.81$e$-02)* & -0.272 (3.38$e$-02)* \\
avgICTF    & 0.045 (7.31$e$-01)* &  -0.067 (6.07$e$-01)*&  -0.084 (5.20$e$-01)*& -0.135 (3.01$e$-01)*& -0.225 (8.14$e$-02)* & -0.183 (1.58$e$-01)* \\
SCQ       & -0.221 (8.65$e$-02)* & -0.248 (5.41$e$-02)* & -0.348 (5.92$e$-03)*& -0.399 (1.44$e$-03)* & -0.327 (1.01$e$-02)* & -0.466 (1.53$e$-04)* \\
nSCQ      & -0.376 (2.80$e$-03)* & -0.492 (5.66$e$-05)* & -0.478 (9.45$e$-05)* & -0.525 (1.40$e$-05)* & -0.525 (1.39$e$-05)* & -0.579 (9.87$e$-07)* \\
maxSCQ    & -0.374 (3.00$e$-03)* & -0.489 (6.37$e$-05)* & -0.458 (2.03$e$-04)*& -0.531 (1.09$e$-05)* & -0.509 (2.85$e$-05)* & -0.583 (8.15$e$-07)*\\
\midrule
collSize  & -0.442 (3.64$e$-04) & \underline{-0.545} (5.53$e$-06) & -0.530 (1.08$e$-05)*  & -0.254 (4.85$e$-02)*  & -0.227 (7.90$e$-02)*  & -0.320 (1.19$e$-02)*  \\
numClusters   & -0.079 (9.68$e$-01)*& -0.005 (5.46$e$-01)*& -0.013 (9.20$e$-01)* & -0.074 (6.29$e$-01)* & 0.063 (5.69$e$-01)* & 0.000 (1.00$e$+00)*\\
collClarity  & -0.033 (7.99$e$-01)*& 0.113 (3.84$e$-01)* & 0.066 (6.09$e$-01)*& 0.006 (9.62$e$-01)* & 0.229 (7.49$e$-02)* & 0.104 (4.23$e$-01)*\\
avgDocSim  &	0.090 (9.47$e$-01)*&  -0.009 (4.93$e$-01)*& -0.128 (3.26$e$-01)*  &  -0.120 (3.58$e$-01)* &	0.084 (5.21$e$-01)*&	-0.147  (2.59$e$-01)*\\
sumDocSim &	\textbf{-0.526} (3.59$e$-04)& -0.442 (1.32$e$-05) &	\underline{-0.568} (1.82$e$-06) & -0.083 (5.24$e$-01)* &	-0.051 (6.98$e$-01)* &	-0.201 (1.21$e$-01)*  \\
maxDocSim  &	-0.232 (2.78$e$-01)*&  -0.141 (7.26$e$-02)* &	-0.345 (6.53$e$-03)*&  -0.274 (3.25$e$-02)* &-0.307 (1.59$e$-02)*&-0.321 (1.16$e$-02)*\\
stdevDocSim & 	0.064 (4.98$e$-01)*&  0.088 (6.24$e$-01)* &	0.267 (3.72$e$-02)* & 0.135 (3.01$e$-01)*& -0.020 (8.81$e$-01)* &0.207 (1.09$e$-01)* \\
\midrule
Proposed & \underline{-0.496} (4.79$e$-05)& \textbf{-0.570} (1.67$e$-06) &  \textbf{-0.584} (7.80$e$-07) & \underline{-0.698} (3.96$e$-10)& \textbf{-0.707} (1.97$e$-10)  & \textbf{-0.720} (6.44$e$-11) \\ 
\midrule
\textit{Rank($\Delta$)} & 2 (0.030) &1 (0.250)    &1 (0.016)   &2 (0.003)  & 1 (0.057)  &1 (0.017)  \\
\bottomrule
\end{tabular}
\end{table*}

\begin{table*}
\small
\centering
\caption{Evaluation of the proposed measure and the baselines using three seed documents. The rest settings are identical with the setting for using two seed documents.}
\label{tab:threeseeds_res}
\begin{tabular}{l|c|c|c|c|c|c}
\toprule
    & \multicolumn{3}{c|}{Spearman's rank correlation coefficient ($p$-value)} &  \multicolumn{3}{c}{Pearson linear correlation coefficient ($p$-value)}    \\
             \midrule
             & BM25 & SDR  & AES  & BM25  & SDR   &AES \\
\midrule
QLen       & 0.202 (1.18$e$-01)* & 0.384 (2.23$e$-03)* &  0.186 (1.51$e$-01)*    & 0.203 (1.17$e$-01)*& 0.377 (2.71$e$-03)* &  0.195 (1.32$e$-01)* \\
avgIDF    & -0.174 (1.81$e$-01)*& -0.122 (3.49$e$-01)* & -0.288 (2.43$e$-02)*    & -0.314 (1.38$e$-02)*& -0.234 (6.94$e$-02)* & -0.362 (4.10$e$-03)*\\
sumIDF  & 0.096 (4.62$e$-01)*& 0.137 (2.92$e$-01)* &  -0.066 (6.14$e$-01)*    & -0.086 (5.10$e$-01)*& 0.080 (5.40$e$-01)*  &  -0.161 (2.16$e$-01)*\\
maxIDF   & -0.365 (3.79$e$-03)*  & -0.230 (7.42$e$-02)* & -0.487 (6.91$e$-05)*& -0.585 (7.55$e$-07)*& -0.438 (4.16$e$-04)* &  -0.631 (5.08$e$-08)*\\
stdevIDF    & 0.395 (1.63$e$-03)* & 0.448 (2.90$e$-04)*  &0.456 (2.25$e$-04)*& \underline{0.690} (7.95$e$-10)*& \underline{0.653} (1.19$e$-08) &  \underline{0.689} (8.34$e$-10) \\
SCS     & -0.058 (6.59$e$-01)*& -0.153 (2.39$e$-01)* & -0.213 (9.91$e$-02)*& -0.208 (1.07$e$-01)*& -0.208 (1.07$e$-01)*&  -0.264 (4.01$e$-02)*\\
avgICTF & 0.061 (6.42$e$-01)*    & -0.058 (6.58$e$-01)*& -0.078 (5.48$e$-01)* & -0.149 (2.52$e$-01)*& -0.216 (9.42$e$-02)* & -0.199 (1.24$e$-01)*\\
SCQ      & -0.203 (1.16$e$-01)*& -0.262 (4.17$e$-02)*& -0.343 (6.85$e$-03)*& -0.390 (1.88$e$-03)*& -0.334 (8.51$e$-03)*&-0.463 (1.74$e$-04)*\\
nSCQ     & -0.369 (3.43$e$-03)*& -0.485 (7.51$e$-05)*& -0.478 (9.72$e$-05)*& -0.526  (1.36$e$-05)*& -0.526 (1.37$e$-05)*&  -0.581 (9.21$e$-07)*\\
maxSCQ   & -0.372 (3.13$e$-03)*& -0.482 (8.38$e$-05)* & -0.465 (1.61$e$-04)*  & -0.533 (9.88$e$-06)* & -0.502 (3.77$e$-05)* & -0.590 (5.50$e$-07)*\\
\midrule
collSize  & -0.442 (3.64$e$-04) & \underline{-0.545} (5.53$e$-06) & -0.530 (1.08$e$-05)*  & -0.254 (4.85$e$-02)*  & -0.227 (7.90$e$-02)*  & -0.320 (1.19$e$-02)*  \\
numClusters   & -0.079 (9.68$e$-01)*& -0.005 (5.46$e$-01)*& -0.013 (9.20$e$-01)* & -0.074 (6.29$e$-01)* & 0.063 (5.69$e$-01)* & 0.000 (1.00$e$+00)*\\
collClarity  & -0.033 (7.99$e$-01)*& 0.113 (3.84$e$-01)* & 0.066 (6.09$e$-01)*& 0.006 (9.62$e$-01)* & 0.229 (7.49$e$-02)* & 0.104 (4.23$e$-01)*\\
avgDocSim  &	0.090 (9.47$e$-01)*&  -0.009 (4.93$e$-01)*& -0.128 (3.26$e$-01)*  &  -0.120 (3.58$e$-01)* &	0.084 (5.21$e$-01)*&	-0.147  (2.59$e$-01)*\\
sumDocSim &	\textbf{-0.526} (3.59$e$-04)& -0.442 (1.32$e$-05) &	\underline{-0.568} (1.82$e$-06) & -0.083 (5.24$e$-01)* &	-0.051 (6.98$e$-01)* &	-0.201 (1.21$e$-01)*  \\
maxDocSim  &	-0.232 (2.78$e$-01)*&  -0.141 (7.26$e$-02)* &	-0.345 (6.53$e$-03)*&  -0.274 (3.25$e$-02)* &-0.307 (1.59$e$-02)*&-0.321 (1.16$e$-02)*\\
stdevDocSim & 	0.064 (4.98$e$-01)*&  0.088 (6.24$e$-01)* &	0.267 (3.72$e$-02)* & 0.135 (3.01$e$-01)*& -0.020 (8.81$e$-01)* &0.207 (1.09$e$-01)* \\
\midrule
Proposed  & \textbf{-0.574} (3.53$e$-05)  & \underline{-0.503} (1.31$e$-06) &  \textbf{-0.572} (1.45$e$-06) & \textbf{-0.709} (1.46$e$-10) & \textbf{-0.710}  (1.62$e$-10) & \textbf{-0.727} (3.23$e$-11)\\
\midrule
\textit{Rank($\Delta$)} & 1 (0.048) &2 (0.042)    &1 (0.004)   &1 (0.019)  & 1 (0.057)  &1 (0.038)  \\
\bottomrule
\end{tabular}
\end{table*}

The proposed measure outperforms the baselines in most cases, and follows the best-performing baseline measure by a small gap in a few cases. While it has the second best estimate for the rank quality estimate of BM25, it outperforms all measures for the rank quality estimate of SDR and AES. Besides, $p$-values in the proposed measure are much smaller than most baselines.

For estimating the relative rank qualities in SRs evaluated by Spearman correlation coefficient, the two measures, sumDocSim, and collSize, show the best performance among baselines, and are comparable to the proposed measure. 
The proposed measure and the two measures utilize the common element: the number of candidate documents.
The strong impact of the number of candidate documents on the rank quality of SR presents that the screening step and the previous retrieval step are closely related. 
As the retrieval step collects the small number of candidate documents with a high precision, the rank quality tends to be higher in the screening step.

Unlike the proposed measure, sumDocSim and collSize do not use example documents. However, we argue that all three measures are proposed for estimating the topic broadness of SRs. Their high performances support our hypothesis that the topic broadness of SRs is a factor that underlies the ranking qualities of SRs.

For estimating the absolute rank qualities of SRs evaluated by Pearson coefficient, either the proposed measure or stdevIDF has the highest performance.
The high stdevIDF value represents that  query terms have various IDF values, and it shows a strong positive correlation with the ranking quality of SR. Recall that a query is set to a relevant document. This result describes that when relevant documents consist of terms with widespread IDF values (\ie informativeness), a SR tends to have the high rank quality.

Next, we discuss the rest baseline measures. In a family of pairwise document similarity measures, sumDocSim, avgDocSim, and maxDocSim incorporate the similarity values as an indicator directly, while stdevDocSim utilizes the distribution of them. 
As shown in Table~\ref{tab:twoseeds_res}, the three measures present a negative correlation the rank quality in most cases.
It shows that when candidate documents are similar to each other, a SR tends to be difficult to achieve the high rank quality. High similarities in candidate documents mean high similarities between relevant and non-relevant documents, and thus it leads to the high ranking difficulty of SR.

numClusters does not provide a meaningful correlation. We look into the results, and find out the majority of SRs has 2 or 3 as the best number of clusters. This result shows that candidate documents are very similar each other, since they share many term overlaps due to Boolean retrieval. At the same time, the document representation using word embeddings is not effective enough to differentiate them in clustering. Incorporating more distinctive document features in document representations can help identify the smaller document clusters.

In Table~\ref{tab:twoseeds_res}, most QPP measures have negative correlations with the rank qualities of SRs.
This is the opposite direction of correlation when they are used for estimating the query performance.
In the query performance estimation, the high value in a QPP measure represents the high query performance.
We believe the different signs of correlations results from the difference in the document collections between the two tasks. Lastly, the opposite behaviors of the QPP measures confirms that the estimation of rank qualities differs from the estimation of query performance, even though they look like similar tasks at first glance.

We conduct the statistic significance test between the proposed measure and baseline measures, similar to~\citep{ICTIR_2013_significance_test_reviewer2,SIGIR_18_significance_test_reviewer2}. We first generate 30 sub-sample groups and each sub-sample group has 30 randomly selected SRs among 63 SRs. We compute correlation coefficients in each subsample group. Then, the statistical significance tests are conducted on 30 coefficient values by the proposed measure and by a baseline measure. The performance of the proposed measure is statistically significant than most baselines, except a few baseline measures. We note that the baseline measures that have superior or comparable performance to the proposed measure (\eg collSize and sumDocSim) are also the newly proposed measures in this work for estimating the topic broadness of SRs.

\paratitle{Multiple Seed Documents.} The proposed measure assumes two relevant documents are available.
We present the averaged estimate over possible relevant document pairs in Table~\ref{tab:twoseeds_res}.
In this section, we examine the effect of the choice of known relevant documents by studying the variance of the values by individual pairs.
A large variance means that the performance of the proposed measure is highly dependent on the choice of the two seed documents. 
In other words, the proposed measure may not show the good performance presented in Table~\ref{tab:twoseeds_res}, depending on given two seeds.
We calculate the variance within each SR. The variances in CLEF 2018 is $0.0003$ on average.
This very small variance represents that the proposed measure can estimate stable values when different seed documents are used.

We study the cases when more than two relevant documents are available and how to incorporate them in the proposed measure.
In this section, we examine the proposed measure when three relevant documents are available. Similar to the evaluation for two available documents, we first generate document triples from relevant documents in a SR. For each triple, the proposed measure is applied to the three document pairs. 
Lastly, we choose the minimum value among the three.
\footnote{Recall that the proposed measure contains a negative sign in Equation~\ref{Equation:proposed_measure}.}
When a SR has broad topic, some of relevant document pairs may not share high similarities.
By selecting the minimum value among the three pairs in a triple, the proposed measure fully utilizes the information given by the three available documents.

We present the evaluation results in Table~\ref{tab:threeseeds_res}.
The overall performance shows the similar patterns as reported in Table~\ref{tab:twoseeds_res} with two example documents.
Since three example documents provide more information about a given SR than two documents, the correlation in Table~\ref{tab:threeseeds_res} are stronger than those in Table~\ref{tab:twoseeds_res}.

Another question about the assumption using two seed documents is: what if there is less than two relevant documents available.
The proposed measure has a limitation for these cases.
However, the existing literature reports that SR researchers typically have a few documents as examples when they conduct a literature survey~\citep{Seed_driven_SDR}.

\paratitle{Ablation study.}
The proposed measure incorporates two types of document similarity: similarity between two seeds ($sim(s_i,s_j)$) and similarity between a seed and a candidate documents ($sim(s_*,c)$). In this section, we study the impact of each type on the performance of the proposed measure. An ablation test is studied by omitting each document similarity from the proposed measure, and their performances are compared to that of the original proposed measure. Table~\ref{tab:ablation_study} shows the evaluation results by Spearman's $\rho$. Both of degraded measures show decreased performance compared to the original measure. It presents both types of document similarity make contribution on the effectiveness of the proposed measure. While the performance of the measure without $sim(s_i,s_j)$ shows a marginal decrease, the performance of the measure without $sim(s_*,c)$ has a large drop. This result suggests that $sim(s_*,c)$ plays a major role in the performance of the proposed measure. Additionally, the large decrease by eliminating $sim(s_*,c)$ also represents the strong impact of the number of candidate documents (\ie the size of document collection) on the ranking quality of SRs, as the proposed measure uses the sum of all $sim(s_*,c)$ in $C$.

\begin{table}
\small
\centering
\caption{The impact of the two types of document similarity in the proposed measure.}
\label{tab:ablation_study}
\begin{tabular}{c|ccc}
\toprule
Spearman's $\rho$& \textbf{BM25} & \textbf{SDR} & \textbf{AES}\\
\midrule
Measure$_{-sim(s,c)}$      & -0.371	& -0.316  &	-0.371 \\
Measure$_{-sim(s,s)}$      & -0.432	& -0.522 & -0.518  \\
\midrule
Proposed measure   & -0.496	& -0.570	&-0.584 \\
\bottomrule
\end{tabular}
\end{table}

%==================================
\section{Conclusion}
\label{sec:Conclusion}
%==================================
We investigate the plausibility of estimating the ranking quality of SR and work towards reducing the manual workload in the screening prioritization. We first make observations on the ranking quality of SRs with different ranking models. Our findings are (1) with different ranking models, the change of ranking quality of SR is very small and (2) when a group of SRs are compared in terms of their ranking qualities, some SRs have the higher ranking qualities that the other SRs and this patterns is consistent across different ranking models. This findings motivate us to hypothesize that there is a defining characteristic in SRs that underpins their ranking quality. To this end, we propose a topic broadness as such characteristic. More specifically, our rational is when a SR has broader topic, it is likely to have lower ranking quality. Through the extensive experiments, we demonstrate that the topic broadness is a simple and effective indicator that predicts the ranking quality of SR. We believe researchers can use the ranking quality signals to anticipate how long the screening will take until they find all relevant documents and eventually to decide when to stop the screening.

\bibliographystyle{ACM-Reference-Format}
\bibliography{references.bib}


\begin{thebibliography}{22}

%%% the Web address.

\ifx \showCODEN    \undefined \def \showCODEN     #1{\unskip}     \fi
\ifx \showDOI      \undefined \def \showDOI       #1{#1}\fi
\ifx \showISBNx    \undefined \def \showISBNx     #1{\unskip}     \fi
\ifx \showISBNxiii \undefined \def \showISBNxiii  #1{\unskip}     \fi
\ifx \showISSN     \undefined \def \showISSN      #1{\unskip}     \fi
\ifx \showLCCN     \undefined \def \showLCCN      #1{\unskip}     \fi
\ifx \shownote     \undefined \def \shownote      #1{#1}          \fi
\ifx \showarticletitle \undefined \def \showarticletitle #1{#1}   \fi
\ifx \showURL      \undefined \def \showURL       {\relax}        \fi
% invisible to TeX
\providecommand\bibfield[2]{#2}
\providecommand\bibinfo[2]{#2}
\providecommand\natexlab[1]{#1}
\providecommand\showeprint[2][]{arXiv:#2}

\bibitem[\protect\citeauthoryear{Butman, Shtok, Kurland, and Carmel}{Butman
  et~al\mbox{.}}{2013}]%
        {ICTIR_2013_significance_test_reviewer2}
\bibfield{author}{\bibinfo{person}{Olga Butman}, \bibinfo{person}{Anna Shtok},
  \bibinfo{person}{Oren Kurland}, {and} \bibinfo{person}{David Carmel}.}
  \bibinfo{year}{2013}\natexlab{}.
\newblock \showarticletitle{Query-performance prediction using minimal
  relevance feedback}. In \bibinfo{booktitle}{\emph{Proceedings of the 2013
  Conference on the Theory of Information Retrieval}}. \bibinfo{pages}{14--21}.
\newblock


\bibitem[\protect\citeauthoryear{Carmel, Yom-Tov, Darlow, and Pelleg}{Carmel
  et~al\mbox{.}}{2006}]%
        {SIGIR2006what_makes_a_query_difficult_four_distances}
\bibfield{author}{\bibinfo{person}{David Carmel}, \bibinfo{person}{Elad
  Yom-Tov}, \bibinfo{person}{Adam Darlow}, {and} \bibinfo{person}{Dan Pelleg}.}
  \bibinfo{year}{2006}\natexlab{}.
\newblock \showarticletitle{What makes a query difficult?}. In
  \bibinfo{booktitle}{\emph{Proceedings of the 29th annual international ACM
  SIGIR conference on Research and development in information retrieval}}.
  \bibinfo{pages}{390--397}.
\newblock


\bibitem[\protect\citeauthoryear{Cronen-Townsend, Zhou, and
  Croft}{Cronen-Townsend et~al\mbox{.}}{2002}]%
        {SIGIR02_clarity}
\bibfield{author}{\bibinfo{person}{Steve Cronen-Townsend}, \bibinfo{person}{Yun
  Zhou}, {and} \bibinfo{person}{W~Bruce Croft}.}
  \bibinfo{year}{2002}\natexlab{}.
\newblock \showarticletitle{Predicting query performance}. In
  \bibinfo{booktitle}{\emph{SIGIR}}. \bibinfo{pages}{299--306}.
\newblock


\bibitem[\protect\citeauthoryear{Hauff, Hiemstra, and de~Jong}{Hauff
  et~al\mbox{.}}{2008}]%
        {preretrievalQPP_survey_CIKM_08}
\bibfield{author}{\bibinfo{person}{Claudia Hauff}, \bibinfo{person}{Djoerd
  Hiemstra}, {and} \bibinfo{person}{Franciska de Jong}.}
  \bibinfo{year}{2008}\natexlab{}.
\newblock \showarticletitle{A survey of pre-retrieval query performance
  predictors}. In \bibinfo{booktitle}{\emph{Proceedings of the 17th ACM
  conference on Information and knowledge management}}. ACM,
  \bibinfo{pages}{1419--1420}.
\newblock


\bibitem[\protect\citeauthoryear{He and Ounis}{He and Ounis}{2006}]%
        {Jornal_SimplifiedClarityScore}
\bibfield{author}{\bibinfo{person}{Ben He} {and} \bibinfo{person}{Iadh Ounis}.}
  \bibinfo{year}{2006}\natexlab{}.
\newblock \showarticletitle{Query performance prediction}.
\newblock \bibinfo{journal}{\emph{Information Systems}} \bibinfo{volume}{31},
  \bibinfo{number}{7} (\bibinfo{year}{2006}), \bibinfo{pages}{585--594}.
\newblock


\bibitem[\protect\citeauthoryear{Jonnalagadda, Goyal, and Huffman}{Jonnalagadda
  et~al\mbox{.}}{2015}]%
        {dataExtract_survey}
\bibfield{author}{\bibinfo{person}{Siddhartha~R Jonnalagadda},
  \bibinfo{person}{Pawan Goyal}, {and} \bibinfo{person}{Mark~D Huffman}.}
  \bibinfo{year}{2015}\natexlab{}.
\newblock \showarticletitle{Automating data extraction in systematic reviews: a
  systematic review}.
\newblock \bibinfo{journal}{\emph{Systematic reviews}} \bibinfo{volume}{4},
  \bibinfo{number}{1} (\bibinfo{year}{2015}), \bibinfo{pages}{78}.
\newblock


\bibitem[\protect\citeauthoryear{Kanoulas, Li, Azzopardi, and Spijker}{Kanoulas
  et~al\mbox{.}}{2017}]%
        {clef_task}
\bibfield{author}{\bibinfo{person}{Evangelos Kanoulas}, \bibinfo{person}{Dan
  Li}, \bibinfo{person}{Leif Azzopardi}, {and} \bibinfo{person}{Rene Spijker}.}
  \bibinfo{year}{2017}\natexlab{}.
\newblock \showarticletitle{CLEF 2017 Technologically Assisted Reviews in
  Empirical Medicine Overview}. In \bibinfo{booktitle}{\emph{CEUR Workshop
  Proceedings}}, Vol.~\bibinfo{volume}{1866}.
\newblock


\bibitem[\protect\citeauthoryear{Kanoulas, Li, Azzopardi, and Spijker}{Kanoulas
  et~al\mbox{.}}{2018}]%
        {CLEF18_overview}
\bibfield{author}{\bibinfo{person}{Evangelos Kanoulas}, \bibinfo{person}{Dan
  Li}, \bibinfo{person}{Leif Azzopardi}, {and} \bibinfo{person}{Rene Spijker}.}
  \bibinfo{year}{2018}\natexlab{}.
\newblock \showarticletitle{{CLEF} 2018 Technologically Assisted Reviews in
  Empirical Medicine Overview}. In \bibinfo{booktitle}{\emph{CLEF}}.
\newblock


\bibitem[\protect\citeauthoryear{Kanoulas, Li, Azzopardi, and Spijker}{Kanoulas
  et~al\mbox{.}}{2019}]%
        {clef_19_overview}
\bibfield{author}{\bibinfo{person}{Evangelos Kanoulas}, \bibinfo{person}{Dan
  Li}, \bibinfo{person}{Leif Azzopardi}, {and} \bibinfo{person}{Rene Spijker}.}
  \bibinfo{year}{2019}\natexlab{}.
\newblock \showarticletitle{{CLEF} 2019 Technology Assisted Reviews in
  Empirical Medicine Overview}. In \bibinfo{booktitle}{\emph{Working Notes of
  {CLEF} 2019}}.
\newblock


\bibitem[\protect\citeauthoryear{Kurland, Raiber, and Shtok}{Kurland
  et~al\mbox{.}}{2012}]%
        {twosides_of_same_coin_QPP_ClusterRank}
\bibfield{author}{\bibinfo{person}{Oren Kurland}, \bibinfo{person}{Fiana
  Raiber}, {and} \bibinfo{person}{Anna Shtok}.}
  \bibinfo{year}{2012}\natexlab{}.
\newblock \showarticletitle{Query-performance prediction and cluster ranking:
  two sides of the same coin}. In \bibinfo{booktitle}{\emph{Proceedings of the
  21st ACM international conference on Information and knowledge management}}.
  ACM, \bibinfo{pages}{2459--2462}.
\newblock


\bibitem[\protect\citeauthoryear{Lee and Sun}{Lee and Sun}{2018}]%
        {Seed_driven_SDR}
\bibfield{author}{\bibinfo{person}{Grace~E. Lee} {and} \bibinfo{person}{Aixin
  Sun}.} \bibinfo{year}{2018}\natexlab{}.
\newblock \showarticletitle{Seed-driven Document Ranking for Systematic Reviews
  in Evidence-Based Medicine}. In \bibinfo{booktitle}{\emph{SIGIR}}.
  \bibinfo{pages}{455--464}.
\newblock


\bibitem[\protect\citeauthoryear{Li and Kanoulas}{Li and Kanoulas}{2020}]%
        {UvAWhenToStop}
\bibfield{author}{\bibinfo{person}{Dan Li} {and} \bibinfo{person}{Evangelos
  Kanoulas}.} \bibinfo{year}{2020}\natexlab{}.
\newblock \showarticletitle{When to Stop Reviewing in Technology-Assisted
  Reviews: Sampling from an Adaptive Distribution to Estimate Residual Relevant
  Documents}.
\newblock \bibinfo{journal}{\emph{TOIS}} \bibinfo{volume}{38},
  \bibinfo{number}{4} (\bibinfo{year}{2020}), \bibinfo{pages}{1--36}.
\newblock


\bibitem[\protect\citeauthoryear{Mikolov, Sutskever, Chen, Corrado, and
  Dean}{Mikolov et~al\mbox{.}}{2013}]%
        {word2vec_original}
\bibfield{author}{\bibinfo{person}{Tomas Mikolov}, \bibinfo{person}{Ilya
  Sutskever}, \bibinfo{person}{Kai Chen}, \bibinfo{person}{Greg~S Corrado},
  {and} \bibinfo{person}{Jeff Dean}.} \bibinfo{year}{2013}\natexlab{}.
\newblock \showarticletitle{Distributed representations of words and phrases
  and their compositionality}. In \bibinfo{booktitle}{\emph{NIPS}}.
  \bibinfo{pages}{3111--3119}.
\newblock


\bibitem[\protect\citeauthoryear{Mizzaro, Mothe, Roitero, and Ullah}{Mizzaro
  et~al\mbox{.}}{2018}]%
        {twosides_of_same_coin_QPP_effectiveness_eval}
\bibfield{author}{\bibinfo{person}{Stefano Mizzaro}, \bibinfo{person}{Josiane
  Mothe}, \bibinfo{person}{Kevin Roitero}, {and} \bibinfo{person}{Md~Zia
  Ullah}.} \bibinfo{year}{2018}\natexlab{}.
\newblock \showarticletitle{Query performance prediction and effectiveness
  evaluation without relevance judgments: Two sides of the same coin}. In
  \bibinfo{booktitle}{\emph{The 41st International ACM SIGIR Conference on
  Research \& Development in Information Retrieval}}. ACM,
  \bibinfo{pages}{1233--1236}.
\newblock


\bibitem[\protect\citeauthoryear{O'Mara-Eves, Thomas, McNaught, Miwa, and
  Ananiadou}{O'Mara-Eves et~al\mbox{.}}{2015}]%
        {SR_survey}
\bibfield{author}{\bibinfo{person}{Alison O'Mara-Eves}, \bibinfo{person}{James
  Thomas}, \bibinfo{person}{John McNaught}, \bibinfo{person}{Makoto Miwa},
  {and} \bibinfo{person}{Sophia Ananiadou}.} \bibinfo{year}{2015}\natexlab{}.
\newblock \showarticletitle{Using text mining for study identification in
  systematic reviews: a systematic review of current approaches}.
\newblock \bibinfo{journal}{\emph{Systematic reviews}} \bibinfo{volume}{4},
  \bibinfo{number}{1} (\bibinfo{year}{2015}), \bibinfo{pages}{5}.
\newblock


\bibitem[\protect\citeauthoryear{Roitman}{Roitman}{2018}]%
        {SIGIR_18_significance_test_reviewer2}
\bibfield{author}{\bibinfo{person}{Haggai Roitman}.}
  \bibinfo{year}{2018}\natexlab{}.
\newblock \showarticletitle{Query performance prediction using passage
  information}. In \bibinfo{booktitle}{\emph{The 41st International ACM SIGIR
  Conference on Research \& Development in Information Retrieval}}.
  \bibinfo{pages}{893--896}.
\newblock


\bibitem[\protect\citeauthoryear{Scells, Azzopardi, Zuccon, and Koopman}{Scells
  et~al\mbox{.}}{2018}]%
        {SIGIR18_QUT_keyword_QPP}
\bibfield{author}{\bibinfo{person}{Harrisen Scells}, \bibinfo{person}{Leif
  Azzopardi}, \bibinfo{person}{Guido Zuccon}, {and} \bibinfo{person}{Bevan
  Koopman}.} \bibinfo{year}{2018}\natexlab{}.
\newblock \showarticletitle{Query Variation Performance Prediction for
  Systematic Reviews}. In \bibinfo{booktitle}{\emph{SIGIR}}.
  \bibinfo{pages}{1089--1092}.
\newblock


\bibitem[\protect\citeauthoryear{Shtok, Kurland, and Carmel}{Shtok
  et~al\mbox{.}}{2009}]%
        {ICIIR09_NQC_query_commitment}
\bibfield{author}{\bibinfo{person}{Anna Shtok}, \bibinfo{person}{Oren Kurland},
  {and} \bibinfo{person}{David Carmel}.} \bibinfo{year}{2009}\natexlab{}.
\newblock \showarticletitle{Predicting query performance by query-drift
  estimation}. In \bibinfo{booktitle}{\emph{ICTIR}}. \bibinfo{pages}{305--312}.
\newblock


\bibitem[\protect\citeauthoryear{Sneyd and Stevenson}{Sneyd and
  Stevenson}{2021}]%
        {SheriffieldWhenToStop}
\bibfield{author}{\bibinfo{person}{Alison Sneyd} {and} \bibinfo{person}{Mark
  Stevenson}.} \bibinfo{year}{2021}\natexlab{}.
\newblock \showarticletitle{Stopping Criteria for Technology Assisted Reviews
  based on Counting Processes}. In \bibinfo{booktitle}{\emph{SIGIR}}.
  \bibinfo{pages}{2293--2297}.
\newblock


\bibitem[\protect\citeauthoryear{Voorhees}{Voorhees}{1985}]%
        {ClusterHypothesis_1985}
\bibfield{author}{\bibinfo{person}{Ellen~M. Voorhees}.}
  \bibinfo{year}{1985}\natexlab{}.
\newblock \showarticletitle{The Cluster Hypothesis Revisited}. In
  \bibinfo{booktitle}{\emph{Proceedings of the 8th Annual International ACM
  SIGIR Conference on Research and Development in Information Retrieval}}
  \emph{(\bibinfo{series}{SIGIR '85})}. \bibinfo{publisher}{ACM},
  \bibinfo{address}{New York, NY, USA}, \bibinfo{pages}{188--196}.
\newblock
\showISBNx{0-89791-159-8}
\urldef\tempurl%
\url{https://doi.org/10.1145/253495.253524}
\showDOI{\tempurl}


\bibitem[\protect\citeauthoryear{Voorhees}{Voorhees}{2004}]%
        {TRECRobust2004overview}
\bibfield{author}{\bibinfo{person}{Ellen~M Voorhees}.}
  \bibinfo{year}{2004}\natexlab{}.
\newblock \showarticletitle{Overview of the TREC 2004 Robust Retrieval Track.}.
  In \bibinfo{booktitle}{\emph{Trec}}.
\newblock


\bibitem[\protect\citeauthoryear{Zhao, Scholer, and Tsegay}{Zhao
  et~al\mbox{.}}{2008}]%
        {ECIR08_Similarity_Query_Collection}
\bibfield{author}{\bibinfo{person}{Ying Zhao}, \bibinfo{person}{Falk Scholer},
  {and} \bibinfo{person}{Yohannes Tsegay}.} \bibinfo{year}{2008}\natexlab{}.
\newblock \showarticletitle{Effective pre-retrieval query performance
  prediction using similarity and variability evidence}. In
  \bibinfo{booktitle}{\emph{ECIR}}. \bibinfo{pages}{52--64}.
\newblock


\end{thebibliography}

\end{document}